\documentclass[aps,prl,twocolumn]{revtex4-1}

\usepackage[usenames,dvipsnames,svgnames,table]{xcolor}
\usepackage{graphicx}
\usepackage[dvipsnames]{xcolor}
\usepackage{amsmath}
\usepackage{blindtext}
\usepackage{printlen}
\usepackage{hyperref}
\usepackage{braket}
\usepackage{todonotes}
\usepackage[pass]{geometry}
\uselengthunit{in}
\usepackage{soul}
\usepackage{CJK}

\begin{document}
\begin{CJK*}{UTF8}{gbsn}

\author{Julius de Hond}
\author{Jinggang Xiang (项晶罡)}
\author{Woo Chang Chung}
\altaffiliation{Present address: ColdQuanta Inc., Boulder, CO, United States of America}
\author{Enid Cruz-Col\'{o}n}
\author{Wenlan Chen}
\altaffiliation{Present address: Department of Physics and State Key Laboratory of Low Dimensional Quantum Physics, Tsinghua University, and Frontier Science Center for Quantum Information, Beijing, 100084, China}
\author{William Cody Burton}
\altaffiliation{Present address: Honeywell Quantum Solutions, Broomfield, CO, United States of America}
\author{Colin Kennedy}
\altaffiliation{Present address: Honeywell Quantum Solutions, Broomfield, CO, United States of America}
\author{Wolfgang Ketterle}
\affiliation{Research Laboratory of Electronics, MIT-Harvard Center for Ultracold Atoms, Department of Physics, Massachusetts Institute of Technology, Cambridge, Massachusetts 02139, USA}

\title{Preparation of the spin-Mott state: a spinful Mott insulator of repulsively bound pairs}

\begin{abstract}
We observe and study a special ground state of bosons with two spin states in an optical lattice: the spin-Mott insulator, a state that consists of repulsively bound pairs which is insulating for both spin and charge transport. Because of the pairing gap created by the interaction anisotropy, it can be prepared with low entropy and can serve as a starting point for adiabatic state preparation. We find that the stability of the spin-Mott state depends on the pairing energy, and observe two qualitatively different decay regimes, one of which exhibits protection by the gap.
\end{abstract}

\maketitle
\end{CJK*}

Mott insulator states of ultracold atoms in optical lattices have played a central role in research ultracold atoms research \cite{Georgescu14, Jaksch05}. Because they are a well-isolated low-entropy state protected by an energy gap, such states have been considered as qubits \cite{Weitenberg11}, as a starting point for adiabatic state preparation \cite{Schachenmayer15, Lubasch11}, and for studies of many-body physics \cite{Bloch08}, in particular quantum magnetism \cite{GarciaRipoll04}. They were used, for instance, in seminal work on Heisenberg spin Hamiltonians \cite{Struck11,dePaz13,Simon11} and as a platform to study Rydberg crystals \cite{Schauss15} and magnetic polarons \cite{Koepsell19}.

When the spin degree of freedom is added to a Mott insulator, it opens up low-lying excitations, and much lower temperatures are needed to reach the ground state. For Mott insulators with occupation $N=1$, the energy scale is set by superexchange, the process by which two spins can be swapped via a virtual intermediate state. This energy scale is often smaller than 1 nK (e.g.\ for rubidium atoms). As a result, magnetically ordered ground states were only observed using fermionic lithium (which due to its low mass has comparatively large tunneling and exchange energies) \cite{Mazurenko17} or using special ramping schemes, see e.g.\ Refs.~\cite{Sun21,Greif15}.

\begin{figure*}
    \centering
    \includegraphics{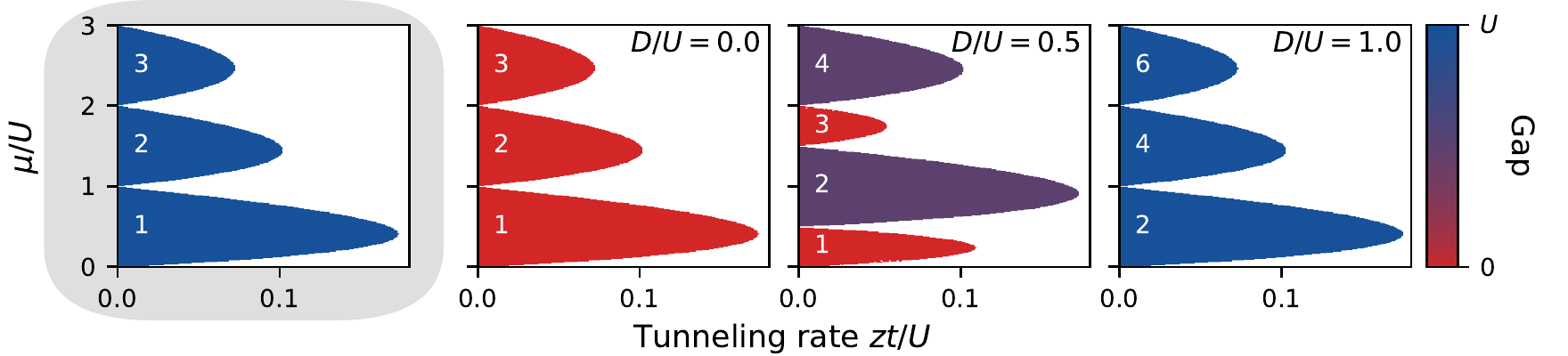}
    \caption{Mean-field phase diagram of the (two-component) Mott insulator showing the number of atoms per site for a particular chemical potential $\mu$ and tunneling rate $t$ (which, in the mean-field model, is enhanced by the coordination number $z$). The leftmost panel contains the familiar lobed phase diagram of the single-component system \cite{Freericks94}. The other panels, from left to right, show the two-component phase diagram for $D/U=0$, $0.5$, and $1.0$, respectively. As $D$ increases, the lobes with an uneven number of particles shrink, to the point where they vanish because the absence of interactions leads to the formation of two independent Mott insulators. The color of the lobes indicates the excitation gap. In the single-component system the first-excited state is a particle-hole pair, which costs an energy $U$ to create. In the two-component system it is a spin excitation with an energy on the order of $D$. The numbers in the lobes indicate $\langle n \rangle = \langle n^a \rangle + \langle n^b \rangle$.}
    \label{fig:phase-diagram}
\end{figure*}

Because preparing spinful ground states is challenging, many experiments probed spin dynamics through quenches, where the initial spin-polarized Mott insulator is suddenly rotated into a spin superposition state. This has enabled study of transport of bound states \cite{Fukuhara13} and spin waves in isotropic \cite{Hild14} and anisotropic \cite{Jepsen20, Jepsen21} $S = 1/2$ Heisenberg models. In the same vein, we recently studied the relaxation of rotated spin states in $S = 1$ Heisenberg models \cite{Chung21}.

Here we show that the situation is drastically different for a spinful Mott insulator with two particles per site. If the on-site interaction energy $U_{AB}$ between opposite spins is considerably lower than that between the same spins $U$, there is an effective pairing energy $D=U-U_{AB}$ favoring the formation of repulsively bound pairs of opposing spins. The ground state of the $N=2$ Mott insulator, then, is a Mott insulator of spin-paired doublons with an excitation gap of $D$. This implies that a spinful $N=2$ Mott insulator has a region in its phase diagram where the excitation gap is of scale $D$ or $U$, which typically corresponds to $50~\mathrm{nK}$ for rubidium, and is thus much larger than the superexchange scale (see Fig.~\ref{fig:phase-diagram}). As a function of $D$, there is a phase transition in the spin domain between a spin superfluid (sometimes referred to as a counterflow superfluid) and a spin insulator. This is in full analogy with the usual superfluid-to-Mott insulator transition in the charge domain \cite{Freericks94}.
The spin-Mott state can serve as an ideal starting point for adiabatic preparation of states with different spin ordering \cite{Schachenmayer15}. One can consider the spin-Mott state for bosons as analogous to the band insulator of fermions for $N=2$ occupation \cite{Lubasch11}, since this state is (in the limit of large pairing energy) a product state of spin-paired doublons on each site.

In this Letter, we demonstrate techniques to prepare and probe the spin-Mott state and study its stability, is protected by the excitation gap if it is sufficiently large.
Our system comprises two different hyperfine states of $^{87}$Rb in a (spin-dependent) optical lattice, which are described by the two-component Bose--Hubbard Hamiltonian \cite{Altman03}. In one dimension and under the assumption of equal tunneling for components $A$ and $B$ this is given by:
\begin{align}
    \label{eq:two-comp-hamiltonian}
    H = &-t \sum_{i} \left( a^\dagger_i a_{i+1} + b^\dagger_i b_{i+1} + \mathrm{H.c.} \right) \notag \\
    & + \frac{U}{2} \sum_{i}\sum_{k\in \left\{a, b\right\}} n_i^k \left( n_i^k - 1 \right) + U_{AB} \sum_i n_i^a n_i^b.
\end{align}
Here $n_i^k$ is the number operator acting on component $k$ on site $i$, $t$ is the nearest-neighbor tunneling parameter, and $U$ and $U_{AB}$ are the intra- and interspecies on-site interactions, respectively, where we have assumed $U = U_{AA} = U_{BB}$.

Restricting ourselves to a deep lattice with a filling of two particles per site, this model maps onto an $S = 1$ Hamiltonian \cite{Altman03, Schachenmayer15}, with the spin-Mott insulator as the simplest ground state for $U_{AB} \ll U$. This is a product state with a single $A$ and $B$ atom per site, and in the spin-1 mapping this corresponds to $|S_z = 0\rangle$.

Correlations become important when 
the pairing energy becomes comparable to the superexchange energy: $D \approx J \equiv -4t^2/U_{AB}$ \cite{Chung21,Kuklov03,Duan03}; in this regime second-order tunneling induces quantum fluctuations of the spin around the spin-Mott insulator. Here the ground state is an $xy$ ferromagnet that contains correlations between sites \cite{Schachenmayer15}; this bears resemblance to the superfluid phase of the single-component system, where the excitation gap vanishes, and where number fluctuations drive correlations between sites \cite{Freericks94}. In the spin model this translates into quantum fluctuations around $|S_z=0\rangle$.

\paragraph*{Experimental setup}
Our experiment starts with a Bose--Einstein condensate of $^{87}$Rb atoms.  A spin mixture of the hyperfine states $A = |F=1, m_F=-1\rangle$ and $B = |F=1, m_F=1\rangle$ is created using a series of microwave sweeps, after which the cloud is loaded into a three-dimensional lattice with depths of at least $25E_R$ to be deep in the Mott-insulating regime. Here $E_R = h^2/2m\lambda^2$ is the recoil energy of a lattice photon with wavelength $\lambda$ for an atom of mass $m$. The interaction between different hyperfine states of rubidium is nearly isotropic \cite{Stamper-Kurn13, Widera06}, and hence for any combination of states $D \approx 0$. The interaction scale $U_{AB}$ can be adjusted, however, by separating the Wannier functions of $A$ and $B$ atoms in the lattice. This can be done using spin-dependent potentials based on the vector AC Stark shift, which, through their effective magnetic field gradient, separate spin states with different magnetic moments. We create such a lattice using a $810$-nm wavelength laser and a tunable polarization gradient \cite{Supplemental}. The two transverse lattices are created using $1064$-nm light.

\begin{figure}
    \centering
    \includegraphics{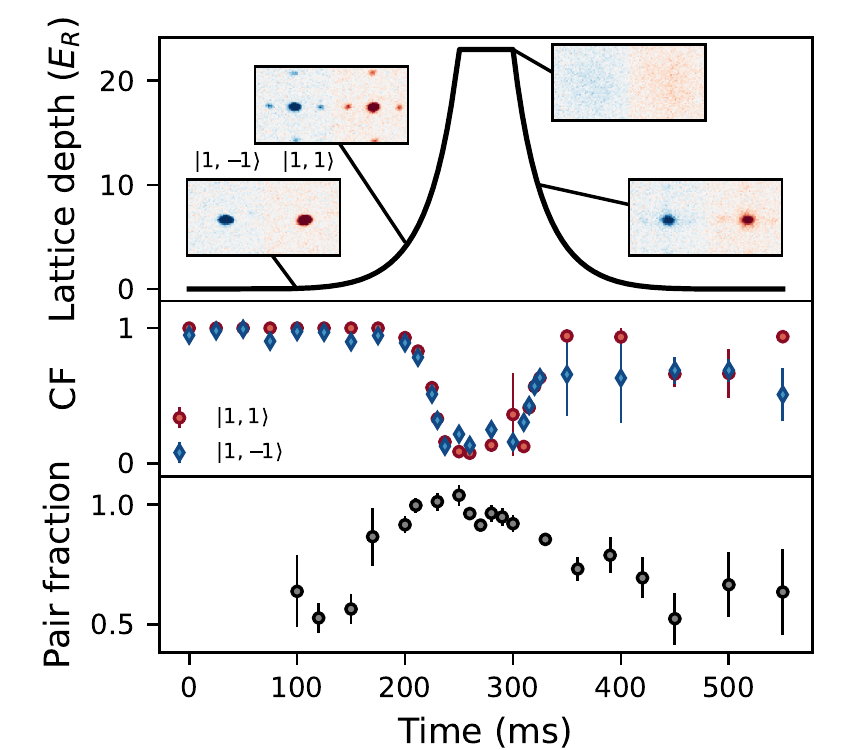}
    \caption{The superfluid-to-Mott insulator phase transition for a system with two spin states.  Starting with a spinor BEC in an equal superposition state of $|1,-1\rangle$ and $|1,1\rangle$, we ramp up the lattice into the Mott insulating regime while $D = U$. In the Mott plateau we observe a dip in condensate fraction (CF), while the pairing fraction approaches unity.}
    \label{fig:sf-to-mi-overview}
\end{figure}

\begin{figure}
    \centering
    \includegraphics{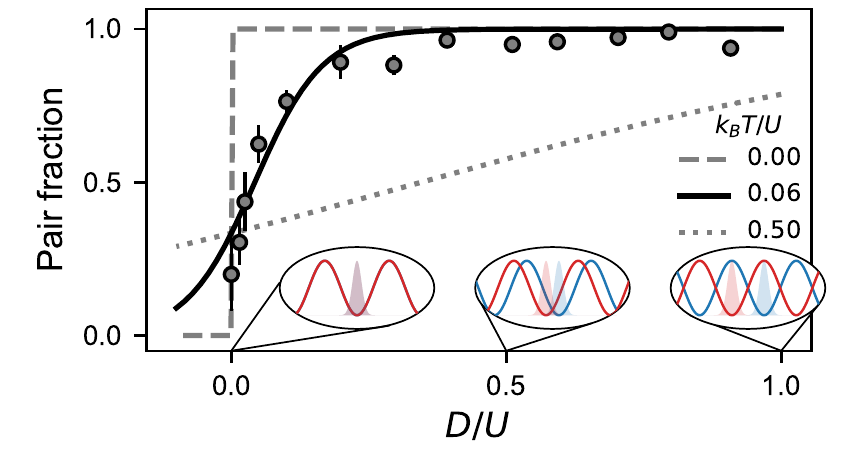}
    \caption{Preparation of the spin-Mott state for various pairing energies $D$.  The decrease of the pair fraction for small $D$ is explained as the effect of finite temperature. Lines are based on the single-site thermal model of Eq.~(\ref{eq:thermal-spdf}) at various temperatures; the insets show the lattice configuration used for various values of $D/U$.}
    \label{fig:loading-spin-mott}
\end{figure}

\paragraph*{Preparing the spin-Mott insulator}
For $D \approx U$, the absence of interspecies interactions leads to the formation of two independent Mott insulators (see Fig.~\ref{fig:phase-diagram}). Here the system exhibits a large excitation gap which we have measured through lattice modulation \cite{Supplemental}. This is similar to the single-component case which has a gap of $U$; hence it is straightforward to prepare the ground state of the Hamiltonian~(\ref{eq:two-comp-hamiltonian}). We do so by creating an equal mixture of the two spin components, followed by a ramp of the lattice while maintaining $D \approx U$. If the atom number is adjusted to fall within the $N = 2$ Mott insulator plateau (but such that it stays shy of the $N = 3$ sector), we prepare a highly ordered spin state with the same wave function on every site.

The pairing fraction is then measured using the pair-resolved doublon detection protocol as described in Ref.~\cite{Chung21}; in short, we overlap the two spin components by quenching to $D = 0$ (see insets in Fig.~\ref{fig:loading-spin-mott}), and take three measurements using absorption imaging: one of the total atom number, one of the atom number after removing all pairs using a Feshbach resonance, and one after selectively removing just the $AB$ pairs using a Feshbach resonance \cite{kaufman2009radio}. The pairing fraction is given by the ratio of differences of these measurements, which makes it susceptible to shot-to-shot number fluctuations. To mitigate this, all the pairing data presented here are obtained as the average of three measurements in each of the three channels.

The spin-Mott insulator shares many properties with its single-component cousin. To highlight this, we have measured the characteristic superfluid-to-Mott insulator phase transition \cite{Greiner02}, while imaging both components individually using Stern--Gerlach separation during time of flight, see Fig.~\ref{fig:sf-to-mi-overview}. From these images, we can determine the condensate fractions in each spin state. Using our pair measurement protocol, we verify that the spin-Mott insulator (realized for deep lattices) has a pairing fraction close to unity.

The gap of the spin-Mott state becomes smaller as we decrease the pairing energy $D$.  We have explored how small $D$ can become before we observe a degradation of the spin-Mott state due to finite temperature or non-adiabatic loading.  Figure~\ref{fig:loading-spin-mott} shows the initial pairing fraction as a function of $D$, after ramping into a deep lattice ($25~E_R$). We find that it is possible to attain high pairing fractions of over 0.9 for a wide range of initial values of $D$. 

According to matrix product state (MPS) calculations, the spin-Mott state is the ground state for $D > 0.05 U$ when $U/t \approx 10$ \cite{Schachenmayer15}, which corresponds to a lattice depth of $8~E_R$. The imperfect pairing fraction observed for $D < 0.2 U$ in a $25~E_R$ lattice can be explained by finite temperature. We can deduce the temperature from a model where tunneling is assumed to be negligible, and hence the Hamiltonian is diagonal on each site in the basis $\left\{|AA\rangle,|AB\rangle,|BB\rangle\right\}$. Generalizing the treatment of a single-component Mott insulator \cite{Gerbier07}, we obtain the pairing fraction (i.e.\ the population in $|AB\rangle$), for a thermal state $|\psi_T\rangle$ at temperature $T$ as
\begin{equation}
    \label{eq:thermal-spdf}
    \left| \langle AB | \psi_T \rangle \right|^2 = \left[1 + 2\exp \left( -D/k_B T \right)\right]^{-1}.
\end{equation}
This expression provides a good description of our data for $k_B T/U \approx 0.06$, which, for $U/2\pi \approx 1500~\mathrm{Hz}$ corresponds to a temperature of 4 -- 5 nK. This is lower than the temperature of the initial Bose--Einstein condensate due to adiabatic cooling during the lattice ramp. Figures~\ref{fig:sf-to-mi-overview} and \ref{fig:loading-spin-mott} represent the main result of this paper:  the successful preparation of the ground state of a spinful $N=2$ bosonic Mott insulator which has not been accomplished before.

\begin{figure}
    \centering
    \includegraphics{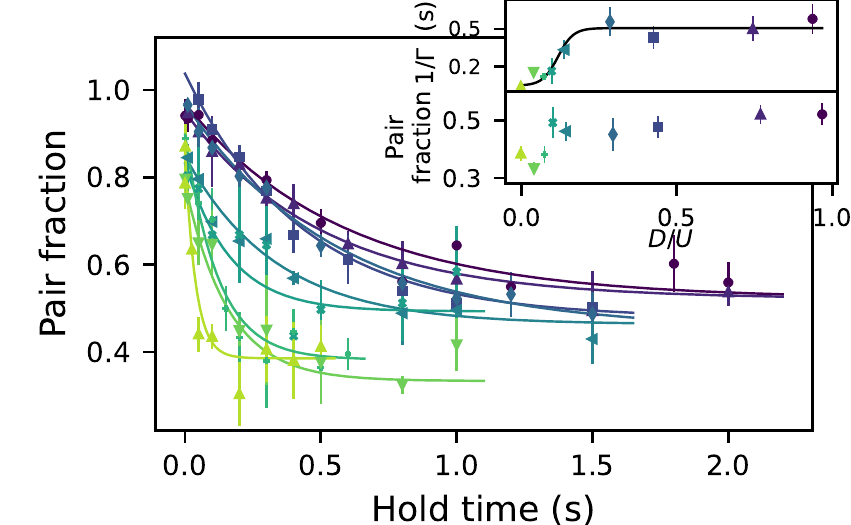}
    \includegraphics{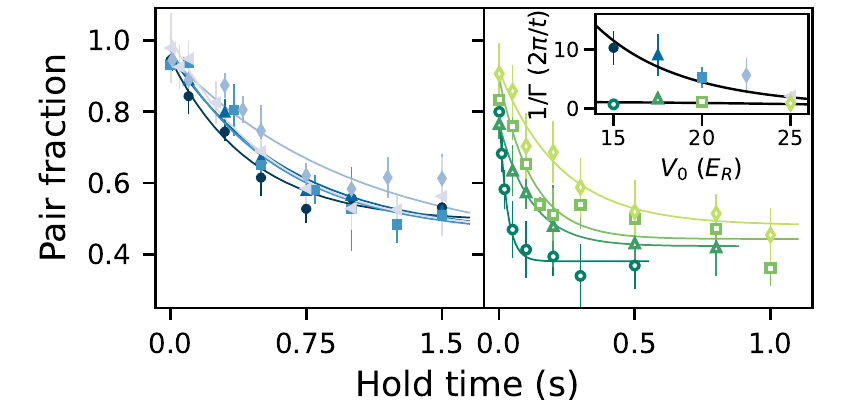}
    \caption{Lifetime of the spin-Mott insulator state. Top: The relaxation shows qualitatively different  behavior as a function of the pairing energy $D$: The  pair fraction either asymptotes towards the infinite temperature limit ($1/3$, if $D$ is small) or not ($1/2$, if $D$ is large). Different colors represent different values of $D/U$. Inset contains fitted lifetimes $1/\Gamma$ and the equi\-li\-bri\-um values of the pair fraction. Bottom: lifetime at various lattice depths, represented by different colors, where $D$ is fixed to be either well within the spin-Mott insulator ($D = 0.3U$, left) or at the isotropic point where $D = 0$ (right). Inset contains fitted lifetime for both data sets scaled by the tunneling time $2\pi/t$. For $D = 0$ we observe that the decay scales with tunneling, whereas in the spin-Mott insulator it is independent of $t$. A single fit of all the lifetimes was done using Eq.~(\ref{eq:lifetime}), which is shown by the black lines.}
    \label{fig:lifetime}
\end{figure}

\paragraph*{Relaxation behavior}
The decay of the spin-Mott state will determine how it can be used as a low-entropy starting point for further experiments. To investigate this, we measure the lifetime as a function of lattice parameters. After the preparation, we ramp $D$ during $100~\mathrm{ms}$ while staying in a deep ($25~E_R$) lattice -- this can be considered a quench since the tunneling rate is on the order of $1~\mathrm{Hz}$. We then lower the lattice to $16~E_R$ and measure how the pairing fraction decays. 

We can distinguish two qualitatively different relaxation regimes as a function of $D$, see the top panel in Fig.~\ref{fig:lifetime}. When quenching $D$ close to $0$, the system quickly approaches the thermal state: an incoherent equal mixture of $|AA\rangle$, $|AB\rangle$, and $|BB\rangle$ on every lattice site, which leads to a pairing fraction of 1/3. For larger values of $D$ the behavior is qualitatively different: not only does the relaxation take longer, the pairing fraction also does not decay to 1/3 over experimentally accessible timescales, rather it goes to $\sim\!1/2$. This is the pairing fraction one would expect if thermalization were constrained to the symmetric subspace; i.e.\ the states $|AB\rangle$ and $\left(|AA\rangle + |BB\rangle\right)/\sqrt{2}$ with each receiving half the population. If all couplings preserve symmetry, the system cannot be taken out of this subspace when starting from the initial state $|AB\rangle$.

The two qualitatively different regimes also show up in the relaxation behavior as a function of lattice depth.  In the bottom panels of Fig.~\ref{fig:lifetime} we compare the system when held in the spin-Mott phase and after quenching to the point where $D=0$.  For small values of $D$, the decay rate scales linearly with the tunneling rate, while in the spin-Mott state the decay rate is independent of lattice depth.

This behavior is captured by modelling the total decay rate as the sum of a background contribution and a term that depends linearly on tunneling but which is suppressed by $D$:
\begin{equation}
\label{eq:lifetime}
    \Gamma\left( t, D \right) = \Gamma_0 + \left(t/2\pi\right) / \left[ 1 + c_1\exp\left(D/c_2\right) \right].
\end{equation}
Here $c_1$ and $c_2$ are fit parameters, and the form is such that $\Gamma\left(t,D\right) \rightarrow \Gamma_0$ if $D$ is large, as it is in the spin-Mott insulator. This expression gives us a quantitative description of the lifetimes measured in Fig.~\ref{fig:lifetime} (see Ref.~\cite{Supplemental}).

We conjecture that a combination of factors leads to this behavior. For $D \approx 0$ we enter the regime where the temperature of the $N = 2$ plateau is sufficiently large to drive deviations from perfect pairing already observed during loading (see Fig.~\ref{fig:loading-spin-mott}). First order tunneling together with imperfections in the Mott insulator will allow entropy transport from the outer regions of the system inwards, which will rapidly increase the spin temperature. This is a plausible explanation for the observed scaling with the tunneling rate.

At higher values of $D$ the excitation gap should protect against relaxation, and it does to the extent that the lifetimes are much longer and are independent of lattice depth. This makes it unlikely that the decay is caused solely by either tunneling or light scattering-related mechanisms. Nevertheless, the interplay between different effects in our experiment is rather complicated; increasing the lattice depth increases light scattering, but it also increases the confinement. With our current setup it is hard to disentangle such mechanisms.  The nature of the slow spin relaxation  could be the slow creation of atoms in excited bands (which are mobile) due to technical noise of the lattice beams, or grazing collisions with background gas atoms.

\paragraph*{Discussion \& conclusions}
While the spin-Mott insulator itself is just a product state, it can be used as a starting point for adiabatically preparing correlated spin states such as the $xy$ ferromagnet \cite{Altman03, Schachenmayer15}. Similar schemes have been proposed for fermions, where the (gapped) band insulator can be used to adiabatically prepare an antiferromagnet \cite{Lubasch11}. In that case the initial product state is stabilized by the bandgap, whereas in our case it is stabilized by the pairing energy $D$.

Adiabatic state preparation requires that the gaps between many-body states are traversed sufficiently slowly; in a Landau--Zener model of avoided crossings, the maximum rate is set by the coupling between states  \cite{Zener32, Landau32, Rubbmark81}. In a deep lattice this scales with second-order tunneling as $\propto t^2/U$ \cite{Kuklov03, Schachenmayer15, Chung21}. Furthermore, coupling between different many-body states scales inversely with the number of sites in a chain. For our present system the superexchange scale  $4t^2/U$ can be boosted to roughly $10~\mathrm{Hz}$ by decreasing the longitudinal lattice depth to $12E_R$. However, this is comparable to some of the decay rates reported in Fig.~\ref{fig:lifetime}. Therefore, some attempts to adiabatically sweep the spin-Mott state into correlated spin states were not successful.

Other atomic species should be more favorable, including cesium, which has a larger fine-structure splitting which makes it possible to create a spin dependent potential at larger detunings with less light scattering. The lanthanides, which feature spin-orbit coupling in the ground state and hence have a vector AC Stark shift for any lattice detuning, are also an attractive alternative.

The future addition of a quantum-gas microscope to our setup \cite{Bakr09} will mitigate some of these issues. With the single-site resolution, experiments can be performed on short chains with definite length which are fully decoupled from surrounding low-density thermal reservoirs.

In conclusion, we have prepared and characterized the spin-Mott state which is the ground state of the two-component Bose--Hubbard model in deep lattices, which can be mapped onto an $S = 1$ Heisenberg Hamiltonian. This state features a large pairing gap, and is a promising platform for adiabatic preparation of magnetic phases and the study of other many-body phenomena. Additionally, since the spin-Mott state is a product state of repulsively bound pairs it offers a way to study pair superfluidity \cite{Menotti10, Daley09}. In the spirit of Ref.~\cite{Volz06}, one could do this by creating a dilute gas of repulsively bound dimers after reducing the harmonic confinement and emptying out the singly-occupied sites.

\begin{acknowledgments}
We thank Yoo Kyung Lee for a critical reading of our manuscript.
We acknowledge support from the NSF through the Center for Ultracold Atoms and Grant No. 1506369, ARO-MURI Non-equilibrium Many-Body Dynamics (Grant No.\ W911NF14-1-0003), AFOSR-MURI Quantum Phases of Matter (Grant No. FA9550-14-1-0035), ONR (Grant No.\ N00014-17-1-2253), and a Vannevar-Bush Faculty Fellowship. W.C.C.\ acknowledges additional support from the Samsung Scholarship. 
\end{acknowledgments}

\bibliography{bib.bib}

\newcounter{comment}
\newcommand{\comment}[2][]{\todo[color=red!100!green!33, #1]{#2}}
\definecolor{myellow}{rgb}{1., 1., 0.6}
\newcommand{\note}[2][]{\todo[color=myellow, #1]{#2}}
\newcommand{\newmat}[1]{\textcolor{red}{#1}}
\newcommand{\rmtxt}[1]{\st{#1}}

%
%
%
%
%

\section*{Supplemental Material}

\subsection*{Experimental considerations}
\begin{figure}
    \centering
    \includegraphics[width=\columnwidth]{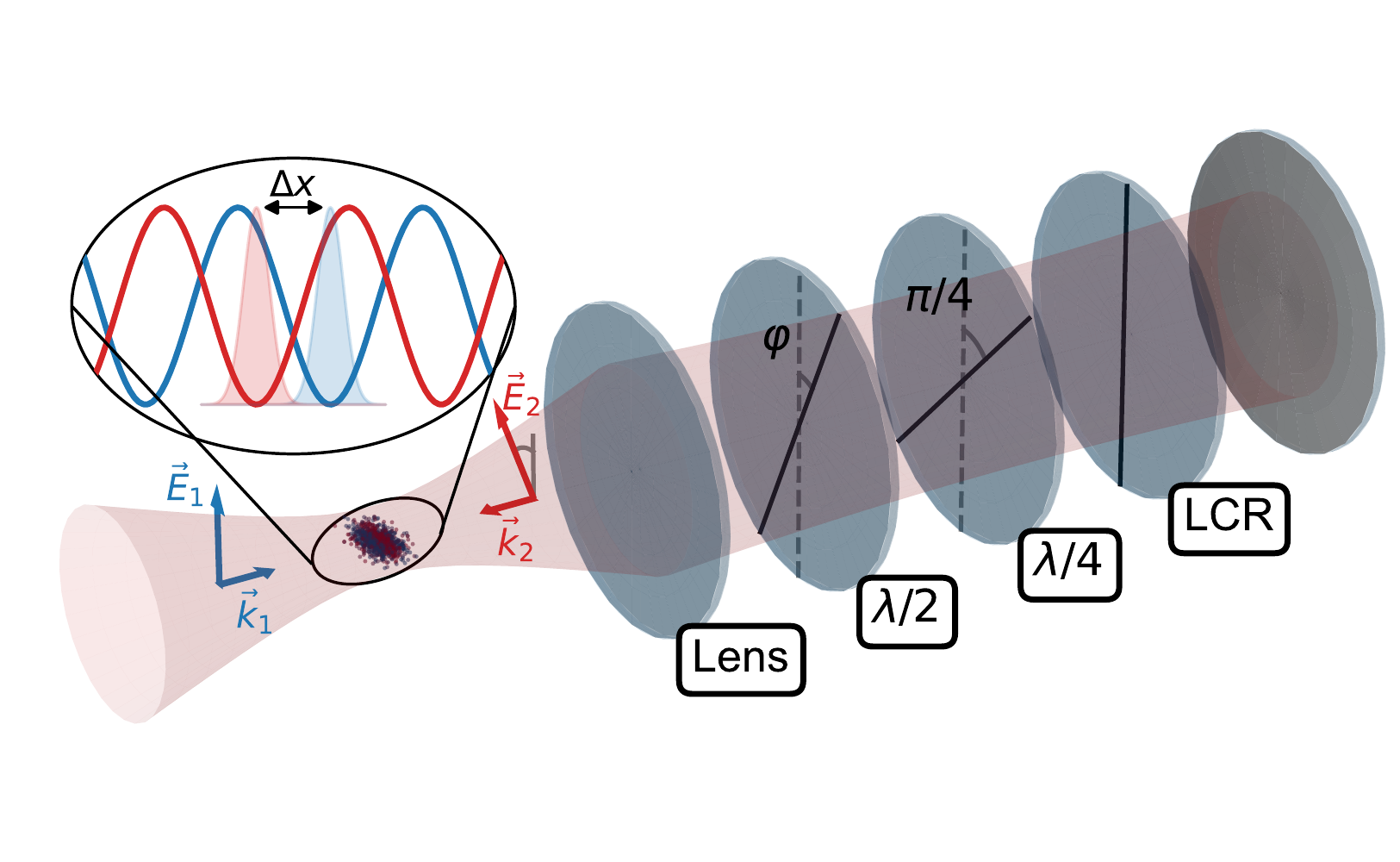}
    \caption{Experimental setup for creating the the spin-dependent lattice. The input beam is reflected through standard wave plates ($\lambda/2$ and $\lambda/4$ signify a half- and quarter-wave plate, respectively) and a liquid-crystal rotator (LCR), which has a tunable retardance. The solid lines denote the slow axes of the polarization components. They rotate the polarization by $\vartheta = 4\varphi + \eta$ in total, where $\eta$ is the retardance of the LCR.}
    \label{fig:cartoon}
\end{figure}

We create a spin-dependent lattice using a $810$-nm wavelength laser. The laser has a detuning from the $D_1$ and $D_2$ lines of $^{87}$Rb that is comparable to the fine-structure splitting, and hence there is an appreciable vector AC Stark shift \cite{Grimm00, LeKien13}. The lattice is generated using a polarization gradient, which is created by rotating the linearly polarized input beam before it is reflected back onto the atoms (see Fig.~\ref{fig:cartoon}) 

The 1D lattice is in a lin-$\vartheta$-lin configuration, where both beams are linearly polarized, and there is an angle $\vartheta$ between them. In general, both the intensity and the polarization gradient contribute to the lattice potential via the scalar and vector AC Stark shift, respectively. At the chosen wavelength of 810 nm, the scalar polarizability is larger than the vector polarizability, and therefore $U_{AB} \approx U$ for small polarization angles.  $U_{AB}$ tends to zero only for near orthogonal polarizations.

\begin{figure}
    \centering
    \includegraphics[width=\columnwidth]{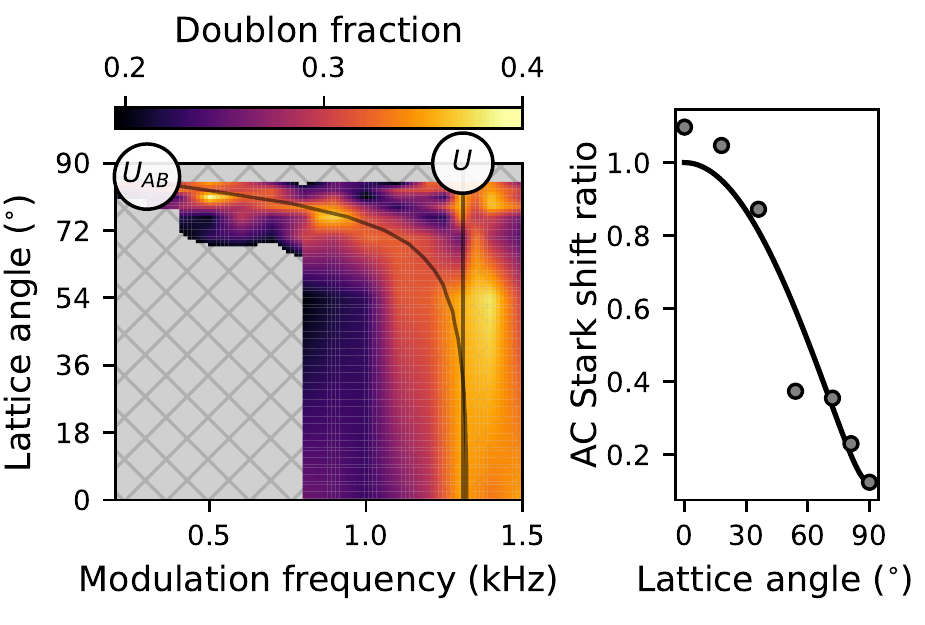}
    \caption{Characterization of the lattice and interaction energies. Left: $U$ and $U_{AB}$ are observed as two separate energy scales through lattice modulation spectroscopy. The measurements were done at different lattice angles, and we interpolated the data in between. The solid lines are calculations based on the overlap of Wannier functions using the lattice separation $\Delta x$ of Eq.~(\ref{eq:interwell-separation}). Right: Measurement of lattice depth for $m_F=1$ atoms using Kapitza--Dirac scattering.  The ratio of lattice depths at $\vartheta = 0^\circ$ and $90^\circ$ is roughly 1/8, as predicted using the lattice wavelength and the fine structure of rubidium.}
    \label{fig:experimental-calibration}
\end{figure}

The optical field in this configuration can be described by a superposition of two circular standing waves.  
For a polarization angle $\vartheta$, the separation between the two circularly polarized sublattices is
\begin{equation}
\label{eq:interwell-separation}
\Delta x / \lambda = \frac{1}{4\pi} \arccos \left( \frac{\cos^2\vartheta - R^2\sin^2\vartheta}{\cos^2\vartheta + R^2\sin^2\vartheta} \right).
\end{equation}
For constant input intensity the lattice depth experienced by $F = 1, |m_F| = 1$ atoms scales with $\vartheta$ as
\begin{equation}
\label{eq:latdepth-scale}
    V_0 \propto \sqrt{\cos^2\vartheta + R^2\sin^2\vartheta}.
\end{equation}
Here $R\approx 1/8$ is the vector-to-scalar AC polarizability ratio for $^{87}$Rb at our wavelength. Using this, $U_{AB}$ can be calculated through the overlap integral of two Wannier functions which have a relative displacement of $\Delta x$. See Fig.~\ref{fig:experimental-calibration} for the numerical result. The scaling of Eq.~(\ref{eq:latdepth-scale}) was confirmed by measuring the lattice depth at various angles through Kapitza--Dirac diffraction \cite{Gould86}, see the right panel in Fig.~\ref{fig:experimental-calibration}.

To calibrate $U_{AB}$, we loaded a small number of atoms into a lattice without any polarization gradient ($\vartheta = 0^\circ$) avoiding any major population of doubly occupied sites.  We then rotated the lattice angle away from zero and lowered the longitudinal lattice depth to $14~E_R$ while the transverse lattices were held at $35~E_R$. Formation of doublons by tunneling was induced by applying a sinusoidal modulation to the lattice depth of $\pm 10\%$. Modulating at a frequency of $U_{AB}/\hbar$ ($U/\hbar$) induces doublon formation between different (the same) spin states \cite{Sias08, Ma11}.

The formed pairs are detected by selectively inducing losses on doubly-occupied sites. Our method is described in detail in Ref.~\cite{Chung21}; briefly, we rotate the lattice back to $\vartheta = 0^\circ$ and transfer the atoms to a pair of states that has a narrow Feshbach resonance at a magnetic field close to $9~\mathrm{G}$ \cite{kaufman2009radio}. By modulating the field around this value, atoms in doubly-occupied sites undergo a loss process. The doublon fraction is obtained by taking the ratio between shots with and without induced losses;  Fig.~\ref{fig:experimental-calibration} shows the spectroscopic measurement of pairing energies.

\subsection*{Fitting the relaxation behavior}
The phenomenological decay rate function of Eq.~(\ref{eq:lifetime}) is inspired by the measurement presented in Fig.~\ref{fig:loading-spin-mott}, where we observe a high-quality spin-Mott state as long as $D$ is above some threshold. Below that threshold thermal effects start to play a role, and we assume that a similar mechanism affects the lifetime:  As the gap shrinks, the system is no longer protected from tunneling from the outer shells of the Mott insulator, and we expect the lifetime to scale with the tunneling rate.

The solid lines shown in Fig.~\ref{fig:lifetime} are based on an overall fit to the lifetime data to Eq.~(\ref{eq:lifetime}) using $c_1 = 0.086$, $c_2 = 0.025U$, and $\Gamma_0 = 2.0~\mathrm{s^{-1}}$.

\subsection*{Mean-field phase diagram}
We have extended the mean-field treatment of the single-component Bose--Hubbard Hamiltonian (see e.g.\ Refs.~\cite{Freericks94,Fisher89}) to the two-component case of Eq.~(\ref{eq:two-comp-hamiltonian}). There are two types of particles annihilated (created) by the operators acting on site $i$: $a_i$ ($a_i^\dagger$) and $b_i$ ($b_i^{\dagger}$). Under the assumption that the correlations between sites are small, we can replace the operators by their deviation from the mean field (e.g.\ by setting $a_i = \langle a \rangle + \delta a_i$ and neglecting cross terms such as $\delta a_i \delta a_j$). This results in the mean-field Hamiltonian:

\begin{align}
    H_\mathrm{MF} = \sum_i &\Big[ -zt \Big( \alpha a_i^\dagger + \alpha^* a_i + \beta b_i^\dagger + \beta^* b_i \notag \\
    &- |\alpha|^2 - |\beta|^2 \Big) + \frac{U}{2} \left[ n^a_i(n_i^a - 1) + n^b_i (n_i^b - 1) \right] \notag \\
    &+ U_{AB} n_i^a n_i^b - \mu (n_i^a + n_i^b) \Big]
\end{align}

Here we have defined $\alpha \equiv \langle a \rangle$ and $\beta \equiv \langle b \rangle$, and use $z$ to denote the number of nearest neighbors. The averages $\alpha$ and $\beta$ take on the role of variational parameters. In principle, they are independent, but in our experiment we strive to achieve similar populations in both components, so by symmetry we can assume they are equal. The mean-field Hamiltonian is defined on the single-site Fock basis $|n^a\rangle\otimes|n^b\rangle$, where $|n^p\rangle \in \left\{ |0\rangle, |1\rangle \cdots |n_\mathrm{max}\rangle \right\}$, and we can obtain the overall ground state by minimizing the lowest-energy eigenstate as a function of $\alpha$.

To obtain the phase diagram, we carry out this procedure for a range of parameters $\left(t/U, \mu/U, U_{AB}/U \right)$. The Mott insulating phase is characterized the absence of number fluctuations, and hence $\langle a \rangle = 0$.
There are two limiting cases; if $U_{AB}/U = 1$ there is effectively no difference between the two components, and everything maps onto the single-component system. If $U_{AB}/U = 0$, on the other hand, the two components do not interact at all, and they form two separable Mott insulators. This is the spin-Mott phase. See Fig.~\ref{fig:phase-diagram} for the phase diagram evaluated at different values of $U_{AB}/U$. As this parameter is tuned away from zero, shells with an uneven integer number of particles develop between the spin-Mott shells that have an even number of particles.

\end{document}